\documentclass[amsthm]{elsart}

\usepackage{yjsco}
\usepackage{natbib}
\usepackage{hyperref}

\usepackage{tikz}
\usetikzlibrary{shapes,arrows}
\usepackage{inputenc}

\usepackage{amsfonts,amsmath,amssymb,amsthm}
\usepackage{verbatim,float}
\usepackage{graphicx,subfigure,url}
\usepackage{dsfont,bm,color,multirow}

\usepackage{listings} 

\newcommand{\code}[1]{\texttt{\lstinline|#1|}}
\let\proglang=\textsf
\newcommand{\pkg}[1]{{\fontseries{b}\selectfont #1}}

\def\R{{\mathbb R}}

\usepackage{Sweave}
\begin{document}

\begin{frontmatter}

\title{Introduction to the \proglang{R} package \pkg{TDA}}

\thanks{Research partially supported by NSF CAREER Grant DMS 1149677.}

\author{Brittany T. Fasy}
\address{Computer Science Department,Tulane University}
\ead{brittany.fasy@alumni.duke.edu}

\author{Jisu Kim}
\address{Department of Statistics, Carnegie Mellon University}
\ead{jisuk1@andrew.cmu.edu}

\author{Fabrizio Lecci}
\address{Department of Statistics, Carnegie Mellon University}
\ead{lecci@cmu.edu}

\author{Cl\'ement Maria}
\address{Geometrica Group, INRIA Sophia Antipolis-M\'editerran\'ee}
\ead{clement.maria@inria.fr}

\begin{abstract}
We present a short tutorial and introduction to using the \proglang{R} package 
\pkg{TDA}, which provides some tools for Topological Data Analysis.
In particular, it includes implementations of functions that, given some data, provide 
topological information about the underlying space, such as the distance function, the distance to a measure,
the kNN density estimator, the kernel density estimator, and the kernel distance.
The salient topological features of the sublevel sets (or superlevel sets) of these functions
can be quantified with persistent homology. We provide an \proglang{R} interface
for the efficient algorithms of the \proglang{C++} libraries \pkg{GUDHI}, \pkg{Dionysus} and \pkg{PHAT}, 
including a function for the persistent homology of the Rips filtration, 
and one for the persistent homology of sublevel
sets (or superlevel sets) of arbitrary functions evaluated over a grid of points.
The significance of the features in the resulting persistence diagrams can be 
analyzed with functions that implement recently developed statistical methods.
The \proglang{R} package \pkg{TDA} also includes the implementation of an algorithm for density 
clustering, which  allows us to identify  the spatial organization of the probability mass associated to 
a density function and visualize it by means of a dendrogram, the cluster tree. 
\end{abstract}


\end{frontmatter}

\section{Introduction}
\label{sec:intro}
Topological Data Analysis (TDA) refers to a collection of methods for finding topological structure in data \citep{carlsson2009topology}. Recent advances in computational topology have made it possible to actually compute topological invariants from data.
The input of these procedures typically takes the form of a point cloud, regarded as possibly noisy observations from an unknown lower-dimensional set $S$ whose interesting topological features were lost during sampling. The output is a collection of data summaries that are used to estimate the topological features of $S$.

One approach to TDA is persistent homology 
\citep{edelsbrunner2010computational}, a method for studying the homology at 
multiple scales simultaneously. More precisely, it provides a framework and 
efficient algorithms to quantify the evolution of the topology of a family of 
nested topological spaces. Given a real-valued function $f$, such as the ones 
described in Section~\ref{sec:distances}, persistent homology describes how the 
topology of the lower level sets $\{x: f(x) \leq t \}$ (or superlevel sets $\{x: 
f(x) \geq t \}$)  change as $t$ increases from~$-\infty$ to~$\infty$ (or 
decreases from $\infty$ to $-\infty$). 
This information is encoded in the persistence diagram,  a multiset of points in 
the plane, each corresponding to the birth-time and death-time of a homological 
feature that exists for some interval of $t$.


This paper is devoted to the presentation of the \proglang{R} package \pkg{TDA},
which provides a user-friendly interface for the efficient algorithms of the 
\proglang{C++} libraries \pkg{GUDHI} \citep{gudhi_stpcoh}, \pkg{Dionysus} \citep{dionysus_C}, and \pkg{PHAT} \citep{phat_C}.
The package can be downloaded from
\url{http://cran.r-project.org/web/packages/TDA/index.html}.

In Section \ref{sec:distances}, we describe how to compute some widely studied 
functions that, starting from a point cloud,  provide topological 
information about the underlying space: the distance function 
(\code{distFct}), the distance to a measure function (\code{dtm}), the k Nearest 
Neighbor density estimator (\code{knnDE}), the kernel density estimator 
(\code{kde}), and the kernel distance (\code{kernelDist}).
Section \ref{sec:persistent} is devoted to the computation of persistence diagrams: the function \code{gridDiag} can be used to compute persistent homology of sublevel sets (or superlevel sets) of functions evaluated over a grid of points; the function \code{ripsDiag} returns the persistence diagram of the Rips filtration built on top of a point cloud. 

One of the key challenges in persistent homology is to find a way to separate the points of the persistence diagram representing the topological noise from the points representing the topological signal.
Statistical methods for persistent homology provide an alternative to its exact
computation. Knowing with high confidence that an approximated persistence
diagram is close to the true--computationally infeasible--diagram is often
enough for practical purposes. 
\cite{fasy2014statistical}, \cite{chazal2014stochastic}, and \cite{chazal2014robust} propose several statistical methods to construct confidence sets for persistence diagrams and other summary functions that allow us to separate topological signal from topological noise.
The methods are implemented in the \pkg{TDA} package, and described in Section 
\ref{sec:persistent}.

Finally, the \pkg{TDA} package provides the implementation of an algorithm for density clustering. 
This method allows us to identify and visualize the spatial organization of the data, without specific knowledge about the data generating mechanism and in particular without any a priori information about the number of clusters. 
In Section \ref{sec:density}, we describe the function \code{clusterTree}, that, given a density estimator, encodes the hierarchy of the connected components of its superlevel sets into a dendrogram, the cluster tree \citep{KpoLux11, kent2013}.

\section{Distance Functions and Density Estimators} 
\label{sec:distances}

As a first example illustrating how to use the \pkg{TDA} package, we show how 
to compute
distance functions and density estimators over a grid.
The setting is the typical one in TDA: a set of points $X=\{x_1, \dots, x_n\} \subset \mathbb{R}^d$ 
has been sampled from some distribution~$P$ and we 
are interested in recovering the topological features of the underlying space
by studying some functions of the data. 
The following code generates a sample of 400 points from the unit circle and constructs
a grid over which we will evaluate the functions.

\begin{Schunk}
\begin{Sinput}
> library("TDA")
> X = circleUnif(400)
> Xlim=c(-1.6, 1.6);   Ylim=c(-1.7, 1.7);   by=0.065
> Xseq=seq(Xlim[1], Xlim[2], by=by)
> Yseq=seq(Ylim[1], Ylim[2], by=by)
> Grid=expand.grid(Xseq,Yseq)
\end{Sinput}
\end{Schunk}

\begin{figure}[h!tb]
\begin{center}
\includegraphics[width=5in]{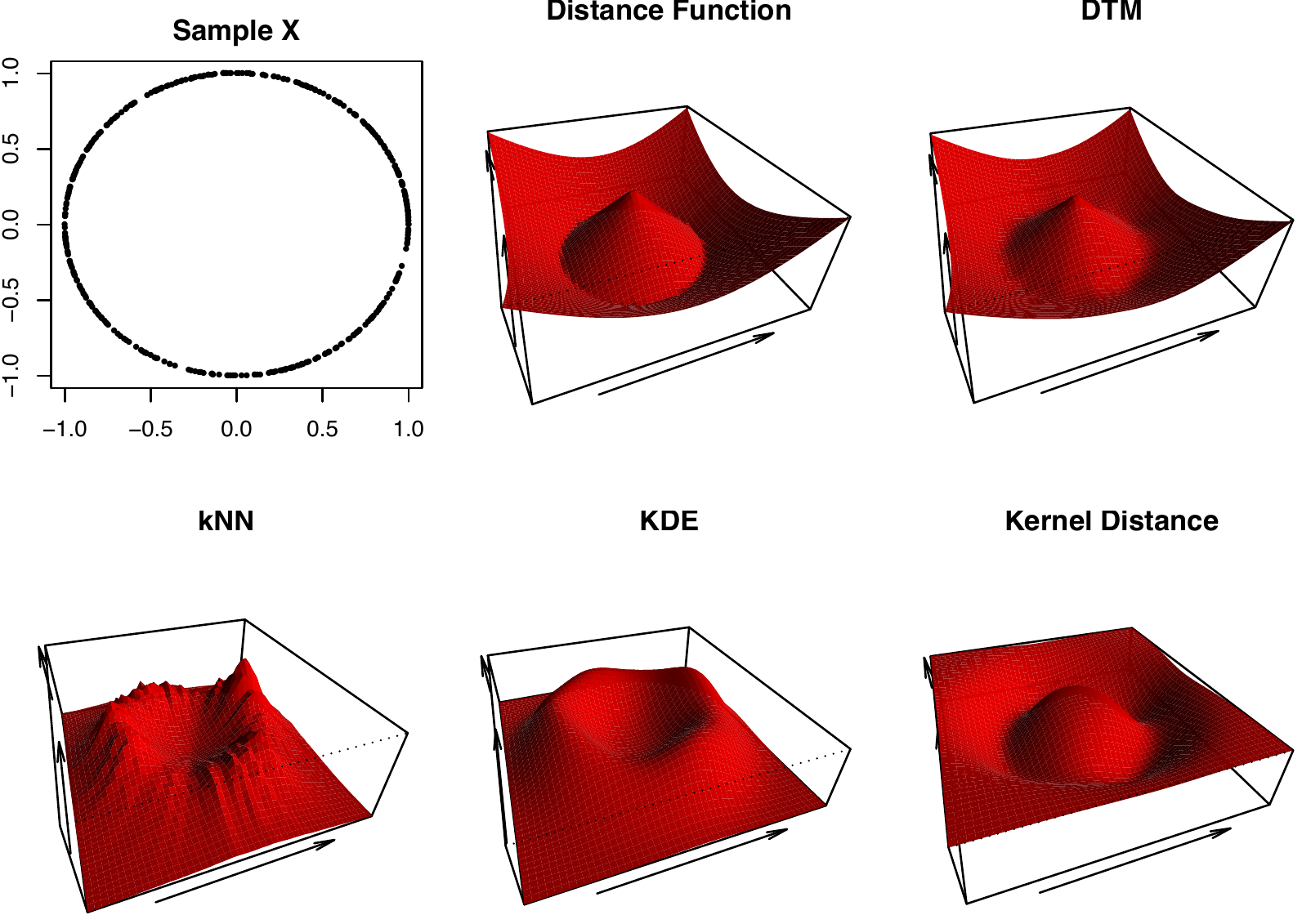}
\end{center}
\caption{distance functions and density estimators evaluated over a grid of points.}
\label{fig:eq8}
\end{figure}

The \pkg{TDA} package provides implementations of the following functions:

\begin{enumerate}
\item The distance function is defined for each $y \in \mathbb{R}^d$ as
$
\Delta(y)= \inf_{x \in X} \| x-y \|_2
$
and is computed for each point of the \code{Grid} with the following code:

\begin{Schunk}
\begin{Sinput}
> distance = distFct(X=X, Grid=Grid)
\end{Sinput}
\end{Schunk}
\item Given a probability measure $P$ on $\R^d$, the distance-to-a-measure 
(DTM) is defined for each $y \in \mathbb{R}^d$ as
$$
d_{m_0}(y) =\sqrt{ \frac{1}{m_0}\int_0^{m_0} ( G_y^{-1}(u))^2 du},
$$
where
$G_y(t) = P( \Vert X-y \Vert \leq t)$ and $0<m_0<1$ is a smoothing parameter. 
The DTM can be seen as a smoothed version of the distance function.
For more details,
see \cite{chazal2011geometric}.

Given $X=\{x_1, \dots, x_n\} \subset \R^d$, the empirical version of the DTM is
$$
\hat d_{m_0}(y) = \sqrt{ \frac{1}{k} \sum_{x_i \in N_k(y)} \Vert x_i-y \Vert^2 },
$$
where $k= \lceil m_0 n \rceil$ and $N_k(y)$ is the 
set containing the $k$ nearest neighbors 
of $y$ among $x_1, \ldots, x_n$.
The DTM is computed for each point of the \code{Grid} with the following~code:

\begin{Schunk}
\begin{Sinput}
> m0=0.1
> DTM=dtm(X=X, Grid=Grid, m0=m0)
\end{Sinput}
\end{Schunk}
\item The $k$ Nearest Neighbor density estimator, for each $y \in \mathbb{R}^d$, is defined as
$$
\hat\delta_k(y)=\frac{k}{n \; v_d \; r_k^d(y)},
$$
where $v_d$ is the volume of the Euclidean $d$-dimensional unit ball and 
$r_k(x)$ is the Euclidean distance form point $x$ to its $k$th closest 
neighbor among the points of $X$.
It is computed for each point of the \code{Grid} with the following code:

\begin{Schunk}
\begin{Sinput}
> k=60
> kNN=knnDE(X=X, Grid=Grid, k=k)
\end{Sinput}
\end{Schunk}
\item The Gaussian Kernel Density Estimator (KDE), for each $y \in \mathbb{R}^d$, is defined as
$$
\hat p_h(y)=\frac{1}{n (\sqrt{2 \pi} h )^d} \sum_{i=1}^n \exp \left( \frac{- \Vert y-x_i \Vert_2^2}{2h^2} \right).
$$
where $h$ is a smoothing parameter.
It is computed for each point of the \code{Grid} with the following code:

\begin{Schunk}
\begin{Sinput}
> h=0.3
> KDE= kde(X=X, Grid=Grid, h=h)
\end{Sinput}
\end{Schunk}
\item The Kernel distance estimator, for each $y \in \mathbb{R}^d$, is defined as
$$
\hat \kappa_h(y)=\sqrt{ \frac{1}{n^2} \sum_{i=1}^n\sum_{j=1}^n K_h(x_i, x_j) + K_h(y,y) - 2 \frac{1}{n} \sum_{i=1}^n K_h(y,x_i)  },
$$
where $K_h(x,y)=\exp\left( \frac{- \Vert x-y \Vert_2^2}{2h^2} \right)$ is the 
Gaussian Kernel with smoothing parameter~$h$.
The Kernel distance is computed for each point of the \code{Grid} with the following code:

\begin{Schunk}
\begin{Sinput}
> h=0.3
> Kdist= kernelDist(X=X, Grid=Grid, h=h)
\end{Sinput}
\end{Schunk}
\end{enumerate}
For this two-dimensional example, we can visualize the functions using 
\code{persp} form the \pkg{graphics} package. For example the following code 
produces the KDE plot in Figure \ref{fig:eq8}:

\begin{Schunk}
\begin{Sinput}
> persp(Xseq,Yseq,matrix(KDE,ncol=length(Yseq), nrow=length(Xseq)),
+ xlab="",ylab="",zlab="",theta=-20,phi=35, ltheta=50, col=2, border=NA, 
+ main="KDE", d=0.5, scale=FALSE, expand=3, shade=0.9)
\end{Sinput}
\end{Schunk}

\subsection{Bootstrap Confidence Bands}
\label{sec:bootstrap}

We can construct a $(1-\alpha)$ confidence band for a function using the bootstrap algorithm, which we briefly describe using the kernel density estimator:

\vspace{0.2cm}

\begin{enumerate}
\item Given a sample $X=\{x_1, \dots, x_n\}$, compute the kernel density estimator $\hat p_h$;
\item Draw $X^*=\{x_1^*, \dots, x_n^*\}$ from $X=\{x_1, \dots, x_n \}$ (with replacement), and compute $\theta^*=\sqrt{n} \Vert \hat p_h^*(x) - \hat p_h(x) \Vert_\infty$, where $\hat p_h^*$ is the density estimator computed using $X^*$;
\item Repeat the previous step $B$ times to obtain $\theta_{1}^*, \dots, \theta_{B}^*$;
\item Compute 
$
q_{\alpha} = \inf \left \{q: \frac{1}{B} \sum_{j=1}^B I(\theta_j^* \geq q) \leq \alpha \right\};
$
\item The $(1-\alpha)$ confidence band for $\mathbb{E}[\hat p_h]$ is
$
\left[\hat p_h - \frac{q_{\alpha}}{\sqrt{n}} \,, \, \hat p_h+ \frac{q_{\alpha}}{\sqrt{n}}\right] .
$
\end{enumerate}

\vspace{0.2cm}

\begin{figure}[h!tb]
\begin{center}
\includegraphics[width=4in]{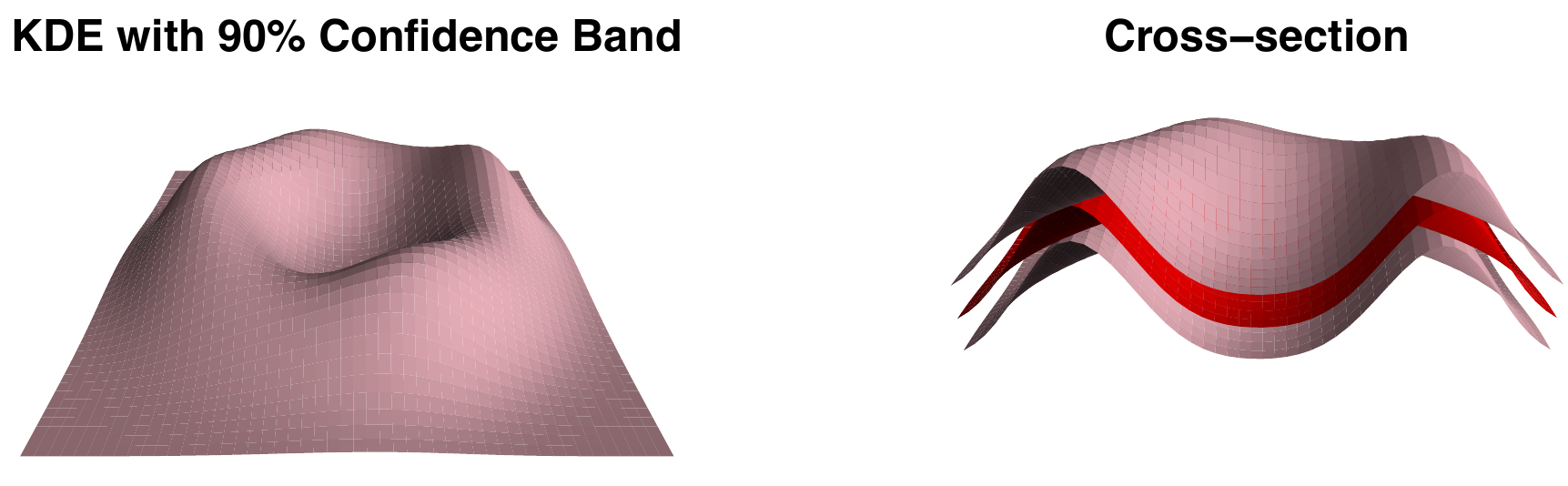}
\end{center}
\caption{the 90\% confidence band for $\mathbb{E}[\hat p_h]$ has the form $[\ell, u]= \left[\hat p_h - q_{\alpha}/\sqrt{n} \,, \, \hat p_h+ q_{\alpha}/\sqrt{n}\right]$. The plot on the right shows a cross-section of the functions: the red surface is the KDE $\hat p_h$; the pink surfaces are $\ell$ and $u$.}
\label{fig:eq10}
\end{figure}

\cite{fasy2014statistical} and \cite{chazal2014robust} prove the validity of the 
bootstrap algorithm for kernel density estimators, distance-to-a-measure, and 
kernel distance, and use it in the framework of persistent homology.
The bootstrap algorithm is implemented in the function \code{bootstrapBand}, which provides the option of parallelizing the algorithm (\code{parallel=TRUE}) using the package \pkg{parallel}.
The following code computes a 90\% confidence band for $\mathbb{E}[\hat p_h]$, showed in Figure \ref{fig:eq10}.

\begin{Schunk}
\begin{Sinput}
> band=bootstrapBand(X=X, FUN=kde, Grid=Grid, B=100, 
+                    parallel=FALSE, alpha=0.1, h=h)
\end{Sinput}
\end{Schunk}

\section{Persistent Homology} 
\label{sec:persistent}

We provide an informal description of the implemented methods of persistent homology. We assume the reader is familiar with the basic concepts and, for a rigorous exposition, we refer to the textbook \cite{edelsbrunner2010computational}.

\subsection{Persistent Homology Over a Grid}
In this section, we describe how to use the \code{gridDiag} function to compute the persistent homology of sublevel (and superlevel) sets of the functions described in Section \ref{sec:distances}. 
The function \code{gridDiag} evaluates a given real valued function over a triangulated grid, constructs a filtration of simplices using the values of the function, and computes the persistent homology of the filtration. From version 1.2, \code{gridDiag} works in arbitrary dimension.
The core of the function is written in \proglang{C++} and the user can choose to compute persistence diagrams using either the \pkg{Dionysus} library or the \pkg{PHAT} library.

The following code computes the persistent homology of the superlevel sets \linebreak (\code{sublevel=FALSE}) of the kernel density estimator (\code{FUN=kde}, \code{h=0.3}) using the point cloud stored in the matrix~\code{X} from the previous example.
The same code would work for the other functions defined in Section \ref{sec:distances} (it is sufficient to replace \code{kde} and its smoothing parameter \code{h} with another function and the corresponding parameter). 
The function \code{gridDiag} returns an object of the class \code{"diagram"}.  
The other inputs are the features of the grid over which the \code{kde} is evaluated (\code{lim} and \code{by}), the smoothing parameter \code{h}, and a logical variable that indicates whether a progress bar should be printed (\code{printProgress}).

\begin{Schunk}
\begin{Sinput}
> Diag = gridDiag( X=X, FUN=kde, h=0.3, lim=cbind(Xlim,Ylim), by=by, 
+                  sublevel=F, library="Dionysus", printProgress=FALSE )$diagram
\end{Sinput}
\end{Schunk}

\begin{figure}[h!tb]
\begin{center}
\includegraphics[width=5in]{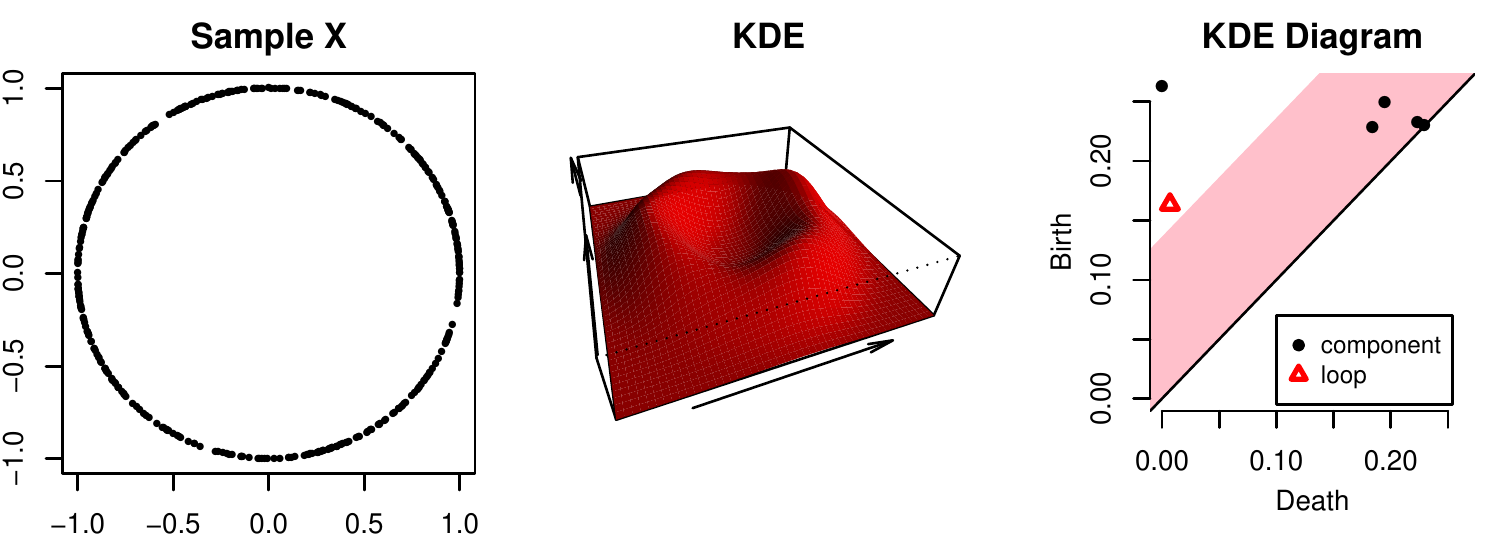}
\end{center}
\caption{The plot on the right shows the persistence diagram of the superlevel sets of the KDE. Black points represent connected components and red triangles represent loops. 
The features are born at high levels of the density and die at lower levels. The pink 90\% confidence band separates significant features from noise.}
\label{fig:eq11b}
\end{figure}

We plot the data and the diagram, using the function \code{plot}, implemented as a standard \code{S3} method for objects of the class \code{"diagram"}. The following command produces the third plot in Figure \ref{fig:eq11b}.

\begin{Schunk}
\begin{Sinput}
> plot(Diag, band=2*band$width, main="KDE Diagram")
\end{Sinput}
\end{Schunk}

The option \code{band=2*band$width} produces a pink confidence band for the persistence diagram, using the confidence band constructed for the corresponding kernel density estimator in the previous section. 
The features above the band can be interpreted as representing significant homological features, while points in the band are not significantly different from noise.
The validity of the bootstrap confidence band for  persistence diagrams of KDE, DTM, and Kernel Distance derive from the \textit{Stability Theorem} \citep{chazal2012structure} and is discussed in detail in \cite{fasy2014statistical} and \cite{chazal2014robust}.

\begin{figure}[h!tb]
\begin{center}
\begin{Schunk}
\begin{Sinput}
> par(mfrow=c(1,2), mai=c(0.8,0.8,0.3,0.1))
> plot(Diag, rotated=T, band=band$width, main="Rotated Diagram")
> legend(0.13,0.25, c("component", "loop"), col=c(1,2), pch=c(20,2), cex=0.9, pt.lwd=2)
> plot(Diag, barcode=T,  main="Barcode")
\end{Sinput}
\end{Schunk}
\includegraphics[width=3.9in]{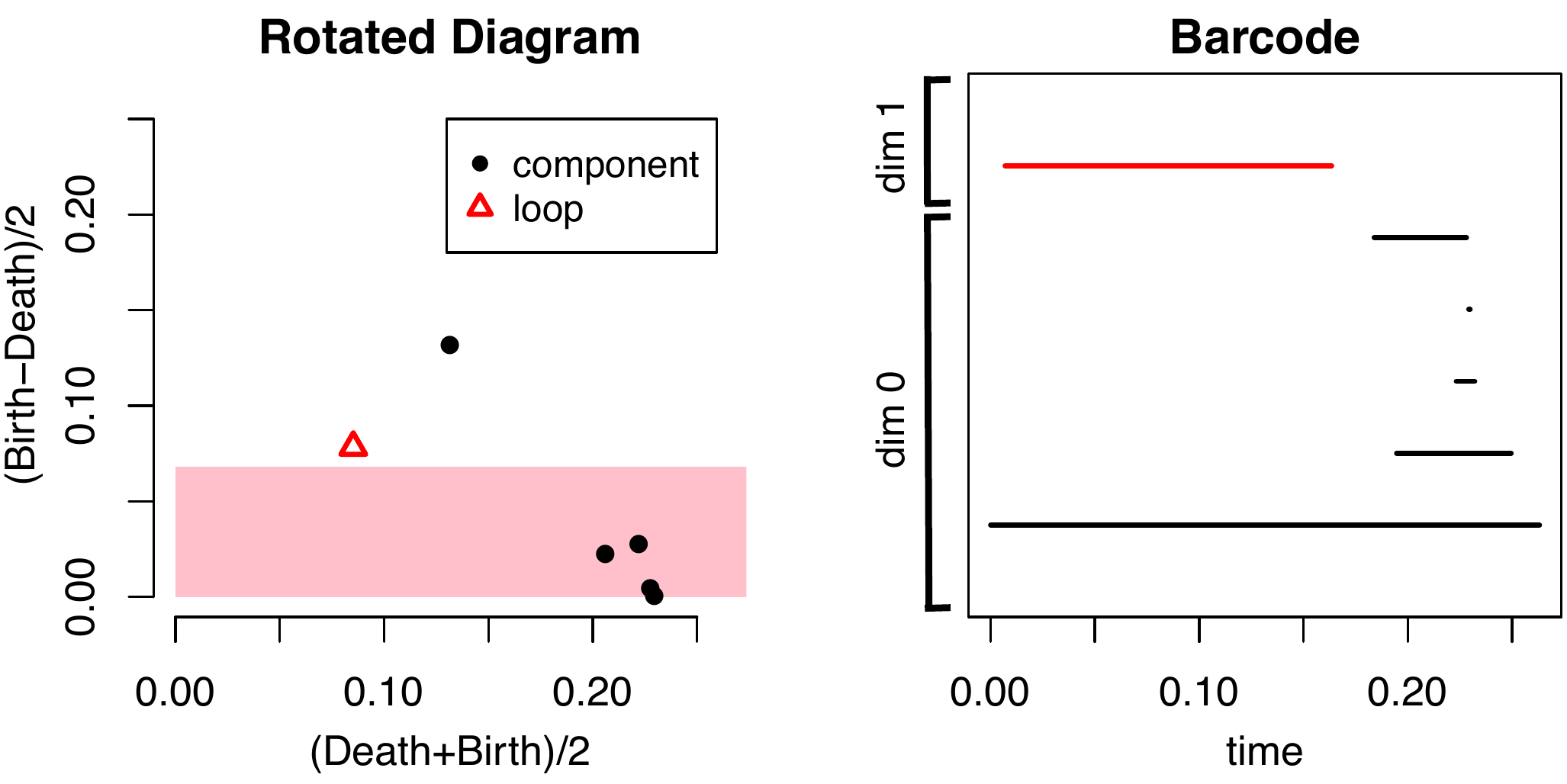}
\end{center}
\caption{Rotated Persistence Diagram and Barcode}
\label{fig:eq11c}
\end{figure}

The function \code{plot} for the class \code{"diagram"} provides the options of rotating the diagram (\code{rotated=TRUE}), drawing the barcode in place of the diagram (\code{barcode=TRUE}), as well as other standard graphical options. See Figure \ref{fig:eq11c}.

\subsection{Rips Diagrams}
The {\em Vietoris-Rips~complex} $R(X,\varepsilon)$
consists of simplices with vertices in \linebreak $X=\{x_1, \dots, x_n\} \subset \mathbb{R}^d$ and diameter
at most $\varepsilon$.  In other words,
a simplex $\sigma$ is included in the complex if each pair of vertices in
$\sigma$ is at most $ \varepsilon$ apart.
The sequence of Rips complexes obtained by gradually
increasing the radius $\varepsilon$ creates a filtration.

The \code{ripsDiag} function computes the persistence diagram of the Rips filtration built on top of a point cloud.
The user can choose to compute the Rips persistence diagram using either the \proglang{C++}  library \pkg{GUDHI}, or \pkg{Dionysus}.
\\The following code generates 60 points from two circles:

\begin{Schunk}
\begin{Sinput}
> Circle1 = circleUnif(60)
> Circle2 = circleUnif(60, r=2) +3
> Circles=rbind(Circle1,Circle2)
\end{Sinput}
\end{Schunk}

\noindent We specify the limit of the Rips filtration and the maximum dimension of 
the homological features we are interested in (0 for components, 1 for loops, 2 
for voids, etc.):

\pagebreak 
\begin{Schunk}
\begin{Sinput}
> maxscale=5           # limit of the filtration
> maxdimension=1       # components and loops
\end{Sinput}
\end{Schunk}

\noindent and we generate the persistence diagram:

\begin{Schunk}
\begin{Sinput}
> Diag=ripsDiag(X=Circles, maxdimension, maxscale,
+               library="GUDHI", printProgress=FALSE)$diagram
\end{Sinput}
\end{Schunk}
Alternatively, using the option \code{dist="arbitrary"} in \code{ripsDiag()}, the input \code{X} can be an $n \times n$ matrix of distances.
This option is useful when the user wants to consider a Rips filtration 
constructed using an arbitrary distance and is currently only available for the 
option \code{library="Dionysus"}.

Finally, we plot the data and the diagram:

\begin{figure}[h!tb]
\begin{center}
\begin{Schunk}
\begin{Sinput}
> par(mfrow=c(1,2), mai=c(0.8,0.8,0.3,0.1))
> plot(Circles, pch=16, xlab="",ylab="", main="Sample")
> plot(Diag, main="Rips Diagram")
> legend(2.5,2, c("component", "loop"), col=c(1,2), pch=c(20,2), cex=0.9, pt.lwd=2)
\end{Sinput}
\end{Schunk}
\includegraphics[width=3.6in]{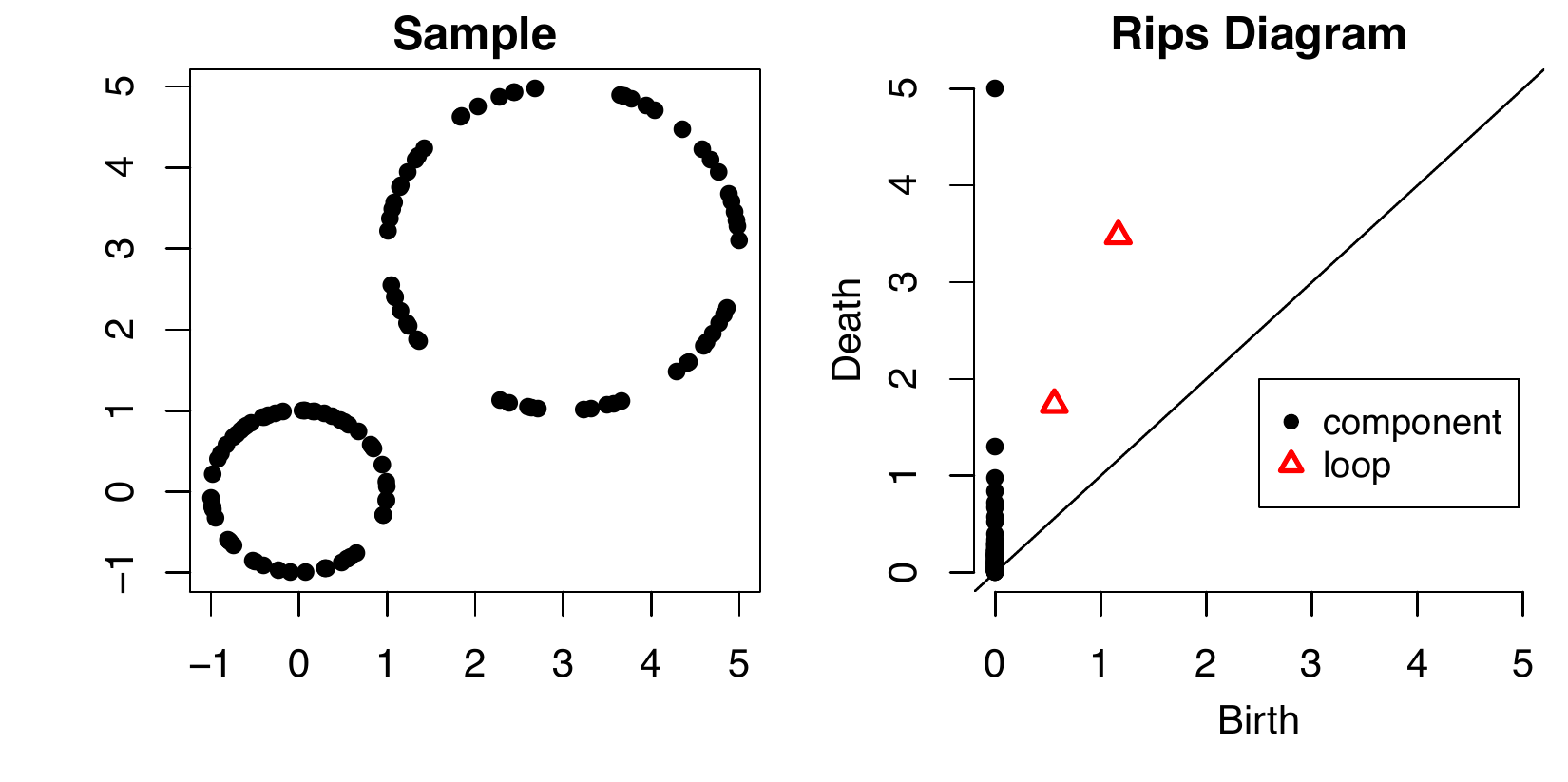}
\end{center}
\caption{Rips persistence diagram. Black points represent connected components and red triangles represent loops.}
\label{fig:eq12d}
\end{figure}

\subsection{Bottleneck and Wasserstein Distances}

Standard metrics for measuring the distance between two persistence diagrams are the bottleneck distance and the $p$th Wasserstein distance \citep{edelsbrunner2010computational}. 
The \pkg{TDA} package includes the functions \code{bottleneck} and 
\code{wasserstein},  which are R wrappers of the \pkg{Dionysus} functions 
``bottleneck\_distance" and ``wasserstein\_distance."

We generate two persistence diagrams of the Rips filtrations built on top of the two (separate) circles of the previous example,

\begin{Schunk}
\begin{Sinput}
> Diag1=ripsDiag(Circle1,maxdimension=1, maxscale=5, printProgress=FALSE)$diagram
> Diag2=ripsDiag(Circle2,maxdimension=1, maxscale=5, printProgress=FALSE)$diagram
\end{Sinput}
\end{Schunk}
and we compute the bottleneck distance and the $2$nd Wasserstein distance between the two diagrams.
In the following code, the option \code{dimension=1} specifies that the 
distances between diagrams are computed using only one-dimensional homological features 
(loops).

\begin{Schunk}
\begin{Sinput}
> print( bottleneck(Diag1, Diag2, dimension=1) )
\end{Sinput}
\begin{Soutput}
[1] 1.15569
\end{Soutput}
\begin{Sinput}
> print( wasserstein(Diag1, Diag2, p=2, dimension=1) )
\end{Sinput}
\begin{Soutput}
[1] 1.680847
\end{Soutput}
\end{Schunk}

\subsection{Landscapes and Silhouettes}

Persistence landscapes and silhouettes are real-valued functions that further summarize the information contained in a persistence diagram. They have been introduced and studied in \cite{bubenik2012statistical}, \cite{chazal2014stochastic}, and \cite{chazal2014subsampling}.
We briefly introduce the two functions.

\begin{figure}[h!tb]
\begin{center}

\includegraphics[width=4.6in]{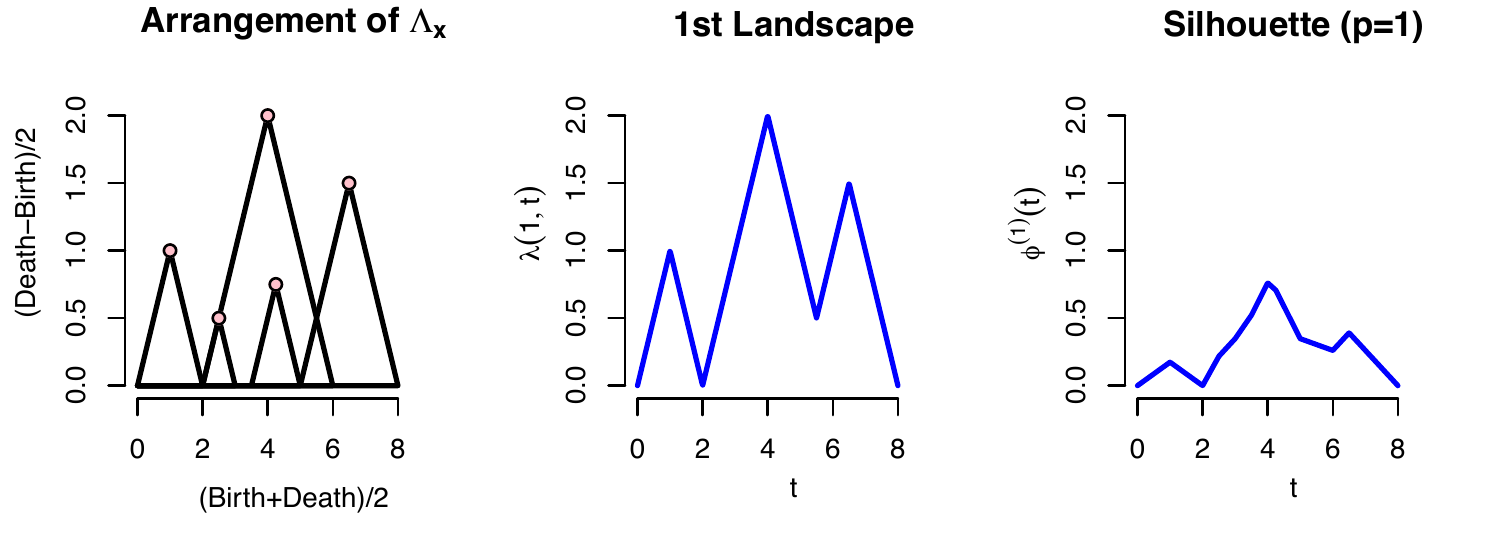}
\end{center}
\caption{	Left: we use the rotated axes to represent a persistence diagram $D$.
	A feature $(b,d) \in D$ is represented by the point
	$(\frac{b+d}{2},\frac{d-b}{2})$ (pink).
	In words,
	the $x$-coordinate is the average parameter value over which the feature
exists,
	and the $y$-coordinate is the half-life of the feature.
        Middle: the blue curve is the landscape $\lambda(1,\cdot)$.
        Right: the blue curve is the silhouette $\phi^{(1)}(\cdot)$.}
\label{fig:eq14a}
\end{figure}

{\bf Landscape.} 
The persistence landscape
is a sequence of continuous, piecewise linear
functions~\mbox{$\lambda \colon  \mathbb{Z}^{+} \times \mathbb{R} \to 
\mathbb{R}$} that provide an
encoding of a persistence diagram.  
To define the landscape, consider the set of functions created by tenting
each point $x=(x_1,x_2)= \left( \frac{b+d}{2}, \frac{d-b}{2}  \right)$
representing a birth-death pair $(b,d)$ in the persistence diagram $D$ as follows:
\begin{equation}\label{eq:triangle}
 \Lambda_p(t) =
 \begin{cases}
  t-x_1+x_2 & t \in [x_1-x_2, x_1] \\
  x_1+x_2-t & t \in (x_1,  x_1+x_2] \\
  0 & \text{otherwise}
 \end{cases}
 =
 \begin{cases}
  t-b & t \in [b, \frac{b+d}{2}] \\
  d-t & t \in (\frac{b+d}{2}, d] \\
  0 & \text{otherwise}.
 \end{cases}
\end{equation}
We obtain
an arrangement of piecewise linear curves by overlaying the graphs of
the functions~\mbox{$\{ \Lambda_x \}_{x}$}; see Figure \ref{fig:eq14a} (left).
The persistence landscape of $D$ is a  summary of this arrangement.
Formally, the persistence landscape of $D$ is the collection of functions
\begin{equation}\label{eq:landscape}
 \lambda(k,t) = \underset{x}{k\text{max}} ~ \Lambda_x(t), \quad
 t \in [0,T], k \in \mathbb{N},
\end{equation}
where $k$max is the $k$th largest value in the set; in particular,
$1$max is the usual maximum~function. see Figure \ref{fig:eq14a} (middle).

{\bf Silhouette.}
Consider a persistence diagram with $N$ off diagonal points $\{(b_j, d_j) \}_{j=1}^N$.
For every $0 < p < \infty$
we define the power-weighted silhouette
$$
\phi^{( p)}(t) = \frac{\sum_{j=1}^N |d_j - b_j|^p \Lambda_j(t)}{\sum_{j=1}^N |d_j - b_j|^p}.
$$

The value $p$ can be thought of as a trade-off parameter between
uniformly treating all pairs in the persistence diagram 
and considering only the most persistent pairs.  Specifically, when $p$ is small,
$\phi^{( p)}(t)$ is dominated by the effect of low persistence features.
Conversely, when $p$ is large,
$\phi^{( p)}(t)$ is dominated by the most persistent features;
see Figure \ref{fig:eq14a} (right).

The landscape and  silhouette functions can be evaluated over a one-dimensional 
grid of points \code{tseq} using the functions \code{landscape} and 
\code{silhouette}. In the following code, we use the persistence diagram from 
Figure \ref{fig:eq12d} to construct the corresponding landscape and silhouette 
for 
one-dimensional features (\code{dimension=1}).  The option \code{KK=1} specifies 
that we are interested in the 1st landscape function, and \code{p=1} is the 
power of the weights in the definition of the silhouette function.

\begin{Schunk}
\begin{Sinput}
> tseq=seq(0,maxscale, length=1000)   #domain
> Land= landscape(Diag, dimension=1, KK=1, tseq)  
> Sil=silhouette(Diag, p=1,  dimension=1, tseq)  
\end{Sinput}
\end{Schunk}

\begin{figure}[h!tb]
\begin{center}
\includegraphics[width=3.6in]{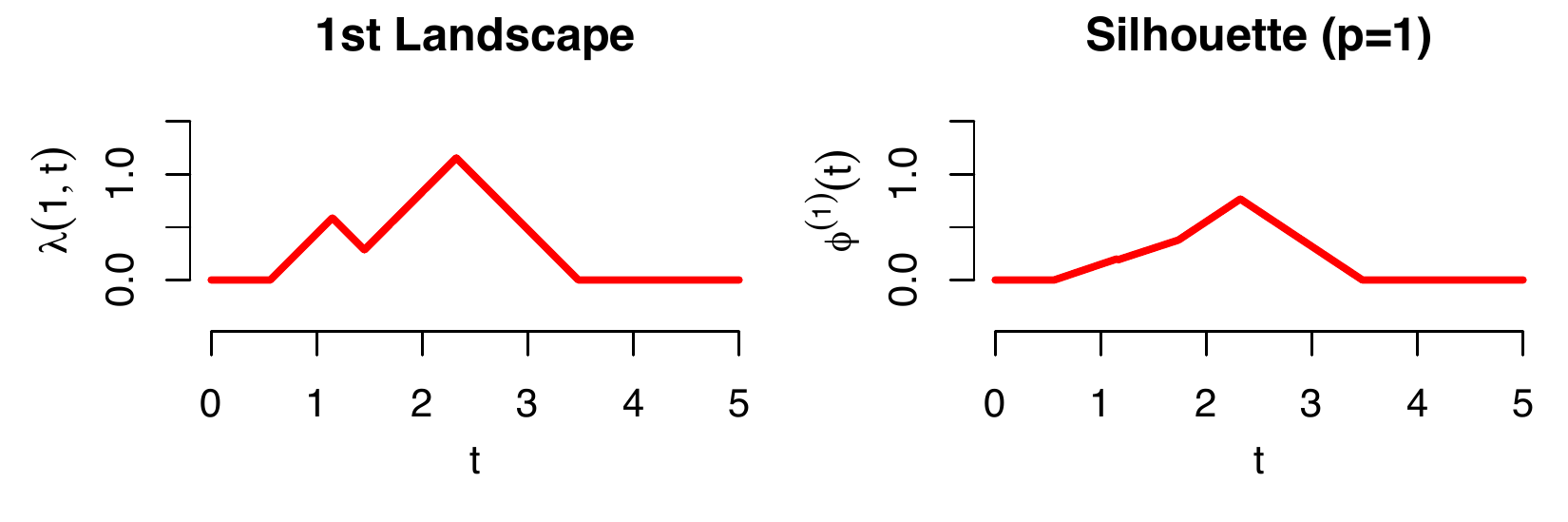}
\end{center}
\caption{Landscape and Silhouette of the one-dimensional features of the diagram of Figure \ref{fig:eq12d}.}
\label{fig:eq14b}
\end{figure}

The functions \code{landscape} and \code{silhouette} return real valued vectors, which can be simply plotted with \code{plot(tseq, Land, type="l"); plot(tseq, Sil, type="l")}. See Figure \ref{fig:eq14b}.

\subsection{Confidence Bands for Landscapes and Silhouettes}

Recent results in \cite{chazal2014stochastic} and \cite{chazal2014subsampling} show how to construct confidence bands for landscapes 
and silhouettes, using a bootstrap algorithm (specifically, the multiplier 
bootstrap).
This strategy is useful in the following scenario.
We have a very large dataset with $N$ points. There is a diagram $D$ and landscape $\lambda$ corresponding to some filtration built on the data. When $N$ is large, computing $D$ is prohibitive. Instead, we draw $n$ subsamples, each of size $m$. We compute a diagram and a landscape for each subsample yielding landscapes $\lambda_1, \dots, \lambda_n $. (Assuming $m$ is much smaller than $N$, these subsamples are essentially independent and identically distributed.) Then we compute $\frac{1}{n}\sum_i \lambda_i$, an estimate of $\mathbb{E}(\lambda_i)$, which can be regarded as an approximation of $\lambda$. The function \code{multipBootstrap} uses the landscapes $\lambda_1, \dots, \lambda_n$ to construct a confidence band for $\mathbb{E}(\lambda_i)$. The same strategy is valid for silhouette functions. We illustrate the method with a simple example.\\
First we sample $N$ points from two circles:

\begin{Schunk}
\begin{Sinput}
> N=4000
> XX1 = circleUnif(N/2)
> XX2 = circleUnif(N/2, r=2) +3
> X=rbind(XX1,XX2)
\end{Sinput}
\end{Schunk}

Then we specify the number of subsamples $n$, the subsample size $m$, and we create the objects that
will store the $n$ diagrams and landscapes:

\begin{Schunk}
\begin{Sinput}
> m=80        # subsample size
> n=10        # we will compute n landscapes using subsamples of size m
> tseq=seq(0,maxscale, length=500)          #domain of landscapes
> Diags=list()                              #here we store n Rips diags
> Lands=matrix(0,nrow=n, ncol=length(tseq)) #here we store n landscapes
\end{Sinput}
\end{Schunk}

For $n$ times, we subsample from the large point cloud, compute $n$ Rips diagrams and the corresponding
1st landscape functions (\code{KK=1}), using one-dimensional features 
(\code{dimension=1}):

\begin{Schunk}
\begin{Sinput}
> for (i in 1:n){
+ 	subX=X[sample(1:N,m),]
+ 	Diags[[i]]=ripsDiag(subX,maxdimension=1, maxscale=5)$diagram
+ 	Lands[i,]=landscape(Diags[[i]], dimension=1, KK=1, tseq )
+ }
\end{Sinput}
\end{Schunk}

Finally we use the $n$ landscapes to construct a 95\% confidence band for the mean landscape

\begin{figure}[h!tb]
\begin{center}
\includegraphics[width=3.5in]{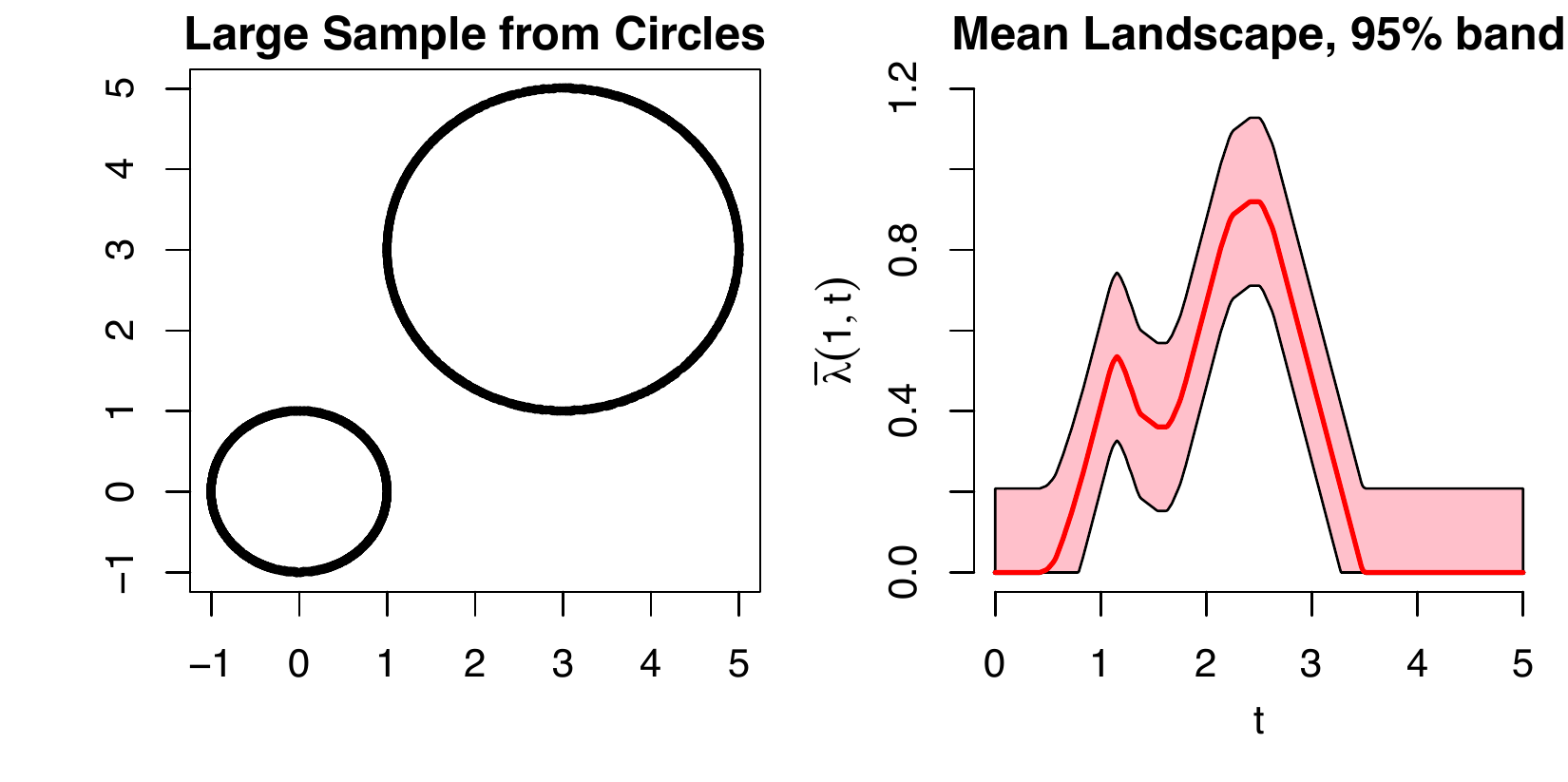}
\end{center}
\caption{95\% confidence band for the mean landscape function.}
\label{fig:eq15f}
\end{figure}

\begin{Schunk}
\begin{Sinput}
> bootLand=multipBootstrap(Lands,B=100,alpha=0.05, parallel=FALSE)
\end{Sinput}
\end{Schunk}
which is plotted by the following code. See Figure \ref{fig:eq15f}.

\begin{Schunk}
\begin{Sinput}
> plot(tseq, bootLand$mean, main="Mean Landscape with 95% band")
> polygon(c(tseq, rev(tseq)), 
+         c(bootLand$band[,1],rev(bootLand$band[,2])), col="pink")
> lines(tseq, bootLand$mean, lwd=2, col=2)
\end{Sinput}
\end{Schunk}

\subsection{Selection of Smoothing Parameters}

An unsolved problem in topological inference is how to
choose the smoothing parameters, for example $h$ for KDE and 
$m_0$ for DTM.

\cite{chazal2014robust} suggest the following method, that we describe here for the kernel density estimator,
but works also for the kernel distance and the distance-to-a-measure.

Let $\ell_1(h),\ell_2(h),\ldots, $
be the lifetimes
of the features of a persistence diagram at scale~$h$.
Let $q_\alpha(h)/\sqrt{n}$
be the width of the confidence band for the kernel density estimator
at scale $h$, as described in Section \ref{sec:bootstrap}.
We define two quantities that measure
the amount of significant information at level $h$:
\begin{itemize}
\item The number of significant features, $N(h) = \# \left\{i:\ \ell(i) > 2\frac{q_\alpha(h)}{\sqrt{n}}\right\}$;
\item The total significant persistence, $S(h) = \sum_i \left[ \ell_i - 2\frac{q_\alpha(h)}{\sqrt{n}}\right]_+$.
\end{itemize}
These measures are small when $h$ is small since $q_\alpha(h)$ is large.
On the other hand,
they are small when $h$ is large since  all of the features of the KDE
smooth out.
Thus we have a kind of topological bias-variance tradeoff.
We choose $h$ to maximize $N(h)$ or~$S(h)$.

The method is implemented in the function \code{maxPersistence}, as illustrated 
in the following example.
First, we sample 1600 points from two circles (plus some clutter noise) and we specify the limits of the grid over which the KDE is evaluated:

\begin{Schunk}
\begin{Sinput}
> XX1 = circleUnif(600)
> XX2 = circleUnif(1000, r=1.5) +2.5
> noise=cbind(runif(80, -2,5),runif(80, -2,5))
> X=rbind(XX1,XX2, noise)
> # Grid limits
> Xlim=c(-2,5)
> Ylim=c(-2,5)
> by=0.2
\end{Sinput}
\end{Schunk}

Then we specify a sequence of smoothing parameters among which we will select the optimal one, the number of bootstrap iterations and the level of the confidence bands to be computed:

\begin{Schunk}
\begin{Sinput}
> parametersKDE=seq(0.1,0.6,by=0.05)
> B=50        # number of bootstrap iterations. Should be large.
> alpha=0.1   # level of the confidence bands
\end{Sinput}
\end{Schunk}

The function \code{maxPersistence} can be parallelized (\code{parallel=TRUE}) and a progress bar can be printed (\code{printProgress=TRUE}):

\begin{Schunk}
\begin{Sinput}
> maxKDE=maxPersistence(kde, parametersKDE, X, lim=cbind(Xlim, Ylim), 
+                    by=by, sublevel=F, B=B, alpha=alpha, 
+                    parallel=TRUE, bandFUN="bootstrapBand")
\end{Sinput}
\end{Schunk}

\begin{figure}[h!]
\begin{center}
\begin{Schunk}
\begin{Sinput}
> par(mfrow=c(1,2), mai=c(0.8,0.8,0.35,0.3))
> plot(X, pch=16, cex=0.5, main="Two Circles + Noise", xlab="", ylab="")
> plot(maxKDE, main="Max Persistence (KDE)")
\end{Sinput}
\end{Schunk}
\includegraphics[width=4.2in]{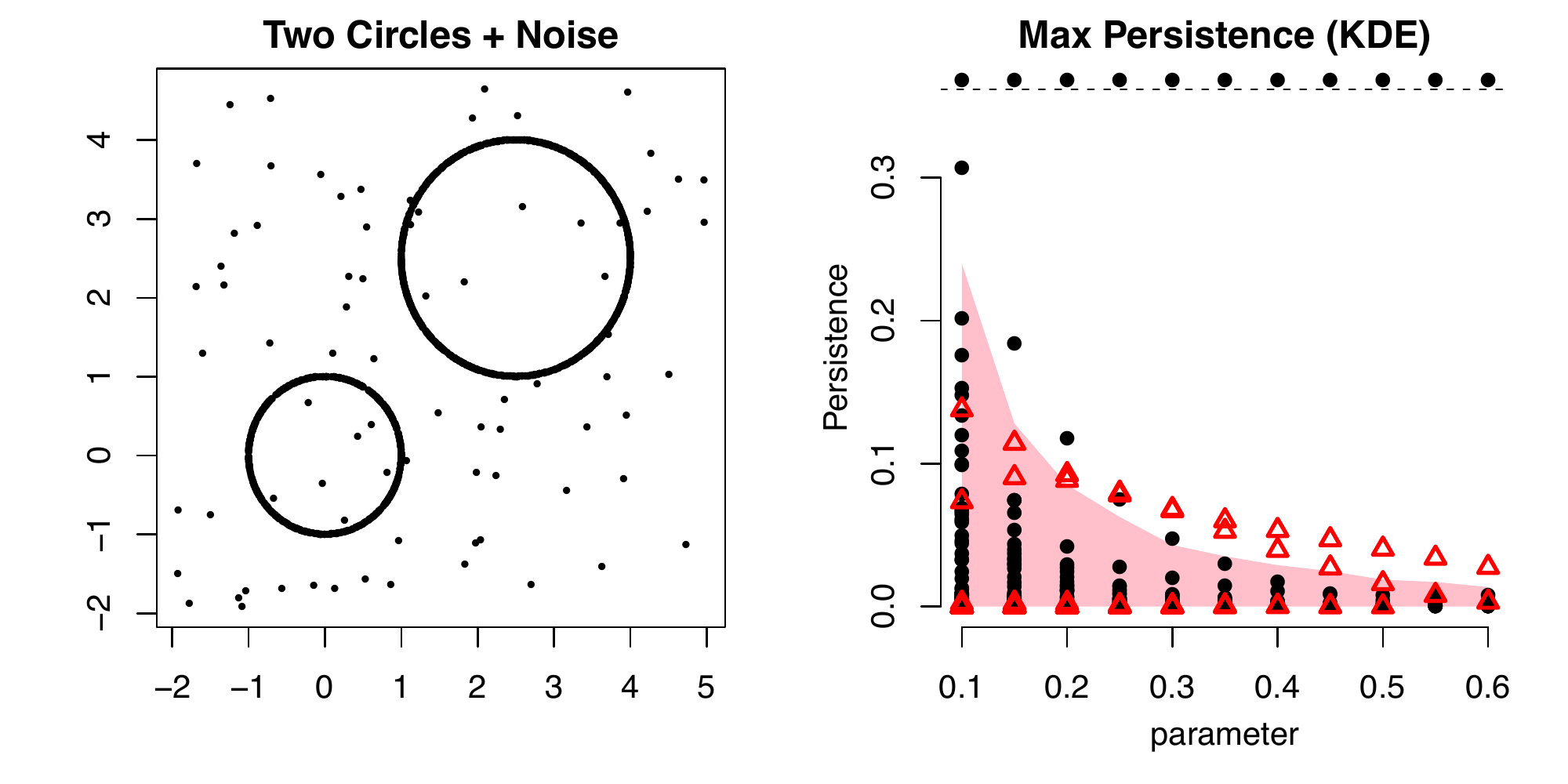}
\end{center}
\caption{Max Persistence Method for the selection of smoothing parameters. For each value of the smoothing parameter we display the persistence of the corresponding homological features, along with a (pink) confidence band that separates the statistically significant features from the topological noise.}
\label{fig:eq16e}
\end{figure}

The \code{S3} methods \code{summary} and \code{plot} are implemented for the class \code{"maxPersistence"}.
We can display the values of the parameters that maximize the two criteria:

\begin{Schunk}
\begin{Sinput}
> print(summary(maxKDE))
\end{Sinput}
\begin{Soutput}
Call: 
maxPersistence(FUN = kde, parameters = parametersKDE, X = X, 
    lim = cbind(Xlim, Ylim), by = by, sublevel = F, B = B, alpha = alpha, 
    bandFUN = "bootstrapBand", parallel = TRUE)

The number of significant features is maximized by 
[1] 0.20 0.25 0.30

The total significant persistence is maximized by 
[1] 0.1
\end{Soutput}
\end{Schunk}
and produce the summary plot of Figure \ref{fig:eq16e}.

\section{Density Clustering}
\label{sec:density}

The final example of this paper demonstrates the use of
the function \code{clusterTree}, 
which is an implementation of Algorithm 1 in \cite{kent2013debacl}.

First, we briefly describe the task of density clustering; we defer the reader to \cite{kent2013} for a more rigorous and complete description.
Let $f$ be the density of the probability distribution~$P$ generating the observed sample $X=\{x_1, \dots, x_n\} \subset \mathbb{R}^d$. For a threshold value $\lambda >0$, the corresponding super level set of $f$ is $L_f (\lambda) := \text{cl} (\{x \in \mathbb{R}^s : f(x) > \lambda\})$, and its $d$-dimensional subsets are called high-density regions. The high-density clusters of $P$ are the maximal connected subsets of $L_f (\lambda)$.
By considering all the level sets simultaneously (from $\lambda=0$ to $\lambda=\infty$), we can record the evolution and the hierarchy of the high-density clusters of $P$.
This naturally leads to the notion of the cluster density tree of~$P$ (see, 
e.g., \cite{hartigan1981consistency}), defined as the collection of sets $T := 
\{L_f (\lambda), \lambda \geq 0\}$, which satisfies the tree property: $A,B \in 
T$ implies that $A \subset B$ or $B \subset A$ or $A \cap B = \emptyset$. We 
will refer to this construction as the $\lambda$-tree. Alternatively, 
\cite{kent2013debacl} introduced the $\alpha$-tree and the $\kappa$-tree, which 
facilitate the interpretation of the tree by precisely encoding the probability 
content of each tree branch rather than the density level. Cluster trees are 
particularly useful for high-dimensional data, whose spatial organization is 
difficult to represent.

\begin{figure}[h!tb]
\begin{center}
\includegraphics[width=4.8in]{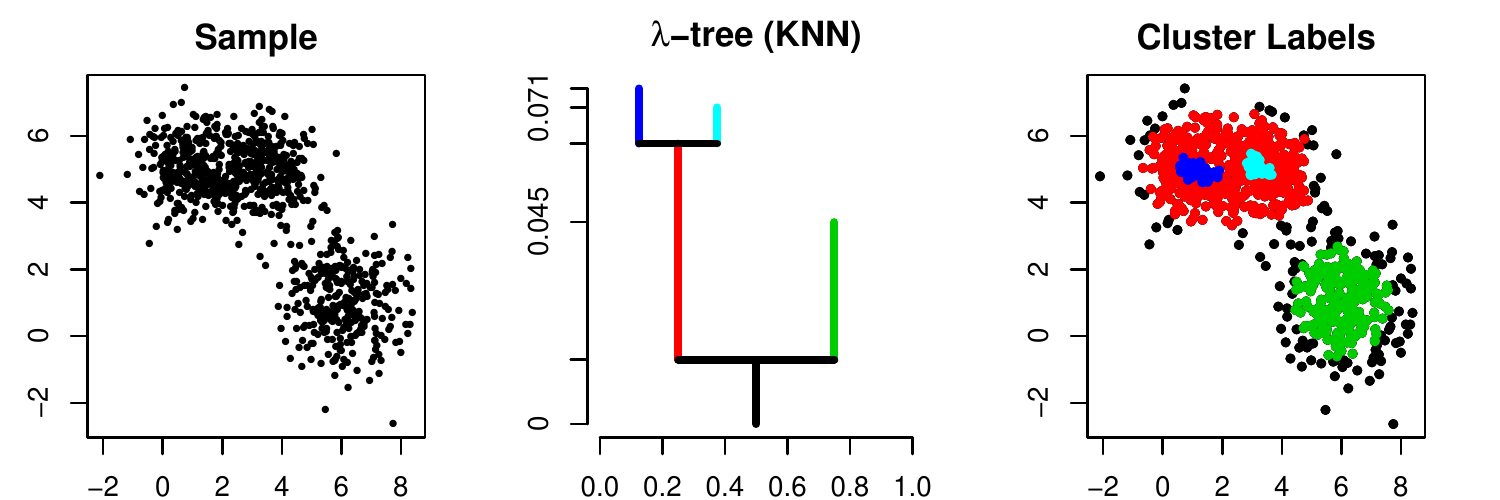}
\end{center}
\caption{The $\lambda$-tree of the k Nearest Neighbor density estimator.}
\label{fig:eq18d}
\end{figure}

We illustrate the strategy with a simple example.
The first step is to generate a 2D point cloud from three (not so well) 
separated clusters (see top left plot of Figure \ref{fig:eq18d}):

\begin{Schunk}
\begin{Sinput}
> X1=cbind(rnorm(300,1,.8),rnorm(300,5,0.8))
> X2=cbind(rnorm(300,3.5,.8),rnorm(300,5,0.8))
> X3=cbind(rnorm(300,6,1),rnorm(300,1,1))
> XX=rbind(X1,X2,X3)
\end{Sinput}
\end{Schunk}

Then, we use the function \code{clusterTree} to compute cluster trees using the 
k Nearest Neighbors density estimator (\code{density="knn"}, \code{k=100} 
nearest neighbors).

\begin{Schunk}
\begin{Sinput}
> Tree=clusterTree(XX, k=100, density="knn")
\end{Sinput}
\end{Schunk}

Alternatively, we can estimate the density using a Gaussian kernel density 
estimator (\code{density="kde"}, \code{h=0.3} for the smoothing parameter). Even 
when the KDE is used to estimate the density, we have to provide the option 
\code{k=100}, so that the algorithm can compute the connected components at each 
level of the density using a k Nearest Neighbors graph.

The \code{"clusterTree"} object \code{Tree} contains information about the $\lambda$-tree, $\alpha$-tree and $\kappa$-tree. The function \code{plot} for objects of the class \code{"clusterTree"} produces the plot in the middle of Figure \ref{fig:eq18d}.

\begin{Schunk}
\begin{Sinput}
> plot(Tree, type="lambda", main="lambda Tree (knn)")
\end{Sinput}
\end{Schunk}

\section{Conclusions}

The \pkg{TDA} package makes topological data analysis accessible to a larger 
audience by providing access to the efficient algorithms for the computation 
of 
persistent homology from the \proglang{C++} libraries \pkg{GUDHI}, 
\pkg{Dionysus}, and \pkg{PHAT} to \proglang{R} users.
Moreover, the package includes the implementation of an algorithm for density 
clustering and a series of tools for the statistical analysis of
persistent homology, including the methods described in
\cite{fasy2014statistical}, \cite{chazal2014stochastic}, and \cite{chazal2014robust}.

We would like \pkg{TDA} to become a comprehensive assistant for the analysis of data from a topological point of view. 
The package will keep providing an \proglang{R} interface for the most efficient libraries for TDA and 
we plan to include new functionalities for all the steps in the data analysis process: new tools for summarizing the topological information contained in the data, methods for the statistical analysis of the summaries, and functions for displaying the results.

\begin{ack}
The authors would  like to thank Fr\'ed\'eric Chazal, Jessi Cisewski, Bertrand 
Michel, Alessandro Rinaldo and Larry Wasserman for their insightful discussions 
and advice.  The authors also thank Dmitriy Morozov for developing  
\pkg{Dionysus} and Ulrich Bauer, Michael Kerber, Jan Reininghaus for 
developing \pkg{PHAT}, and for making these packages freely available.
\end{ack}

\bibliographystyle{elsart-harv}
\bibliography{biblio}  

\end{document}